\documentclass{article}
\usepackage{color}
\usepackage{amssymb}
\usepackage{amsmath}
\usepackage{latexsym}
\usepackage{slashed, amsthm}

\title{Non-compact manifolds with Killing spinors}
\author{C. ~Rugina$^{ \ 1,2}$, A. Ludu$^3$
	\\
	\\ \small{1. Department of Theoretical Physics, IFIN-HH, Magurele, Romania,}
	\\ \small{2. Department of Physics, University of Bucharest, Bucharest, Romania}
	\\ \small{3. Department of Mathematics, Embry-Riddle Aeronautical University, Daytona Beach, FL, USA}
	\\}

\begin{document}

\maketitle 
{\let\thefootnote\relax\footnotetext{{\em Emails}:
		christina.rugina11@alumni.imperial.ac.uk, ludua@erau.edu}}

\begin{abstract}

\noindent
We present a result for non-compact manifolds with invertible Dirac operator, where we link the presence of a massless Killing spinor, with a harmonic, closed conformal Killing-Yano tensor, if one exists for the specific manifold. A couple of examples are introduced.
\end{abstract}

\vskip0.7cm

\vskip0.7cm


\section{Introduction}

\noindent
Killing-Yano (KY) tensors have been studied since 1950, when the KY were firstly introduced \cite{yano}. In general the presence of Killing-Yano tensors indicates also the presence of ancillary St\"{a}ckel-Killing (SK) tensors, hence more hidden symmetries \cite{houri_yasui}. In the literature (\cite{kubi} and references therein) they study the principal conformal Killing-Yano (PCKY) tensor, a special tensor which  builds towers of KY and SK tensors. This means that if a PCKY (which is a closed tensor) is proved to exist in a specific spacetime, then SK and KY tensors can be constructed. SK tensors are symmetrical and help separate in various backgrounds the Hamilton-Jacobi and Klein-Gordon equations, while the KY tensors are antisymmetric and separate the Dirac equation. So these tensors are very useful quantities, which help integrate complicated equations in complicated background metrics. As far as for spinors, it was proved by Lichnerowicz that for compact spaces of positive curvature one can invert the Dirac operator, so solutions to the Dirac equation exist in that case \cite{C52}. In our case, we work with non-compact spin manifolds, for which solutions to the Dirac equation exist.

\medskip
\noindent
We investigate here the relation between conformal KY (CKY) tensors and Killing spinors for non-compact manifolds. It is a well known fact that compact spin manifolds of dimension $\ge 3$ admit a Killing spinor \cite{lichnerowicz} (for a compact manifold the eigenspinors of the Dirac equation corresponding to a lower bound involving the scalar curvature are Killing spinors).
Killing spinors are known in the literature for classifying supergravities e.g. \cite{Y21} and they are both solutions of the Dirac equation and twistor equation \cite{C47}. In 10 and 11 dimensions they classify backgrounds in IIB and M-theory via 5 \cite{Y21}, respectively 6 types \cite{Y22} of Killing spinors.

\medskip
\noindent
In general, when a Killing spinor exists, one can construct a Killing-Yano tensor \cite{cariglia} and hence  a CKY tensor, but to our knowledge in the literature there is not yet a proof on the reciprocal, that is the existence of 
a CKY tensor doesn't necessarily imply the existence of a Killing spinor in that spacetime. A couple of 8-dimensional non-compact Spin(7) holonomy manifolds are introduced.

\medskip
\noindent
There is a lot of work in the literature somewhat mirroring our work here, for instance on symmetry operators that help solve complicated differential equations. In \cite{Benn_Charlton} symmetry operators for the massless Dirac equation for all dimensions and signatures were constructed from arbitrary degree CKY tensors. Also in \cite{Ertem} symmetry operators of the Killing spinor equations were derived from KY tensors and these operators close into an algebra in $AdS_5$. Moreover, a classification of all complete simply-connected Riemannian manifolds with real Killing spinors is given in \cite{Bar}.

\section{Closed, harmonic conformal Killing-Yano tensors imply Killing spinors for certain non-compact manifolds}

\noindent 
Here we present the main result of this paper for non-compact spacetimes (the result holds for compact spaces, too, but this is a known fact). There is a good amount of work done in the literature on determining Killing spinors on negative curvature manifolds, namely $AdS_5$ \cite{C42, C43}, and also work done on non-compact, complete, positive curvature manifolds to further understand the Dirac operator on these manifolds \cite{C50, C51}. We place our work in this diverse and intense context.

\medskip
\noindent
We present the sufficient conditions for the existence of Killing spinor corresponding to an eigenvalue zero Dirac spinor on a non-compact manifold. There are two conditions. The first condition requests the existence of an invertible Dirac operator. The second condition requests the presence of a closed, harmonic conformal Killing-Yano tensor on this manifold.

\medskip
\noindent
We work with non-compact space for which solutions to the Dirac equations exist. In addition, let us assume that a $p$-form $\omega$ exists, and that it is harmonic, closed conformal Killing-Yano tensor.

\medskip
\noindent
We shall now focus for simplicity on a 1-form $\omega$, but the same procedure can be generalized to the arbitrary $p$-forms case by mathematical induction. Since $\omega$ exists, we want to prove that it can be written as below in Eq. (\ref{def_omega}) where $\psi$ is a Killing spinor: 
\begin{equation}\label{def_omega}
\omega_i = \bar{\psi} \gamma_i \psi.
\end{equation}

\section{Construction of a solution for the massless Dirac equation}

\noindent
To prove Eq. (\ref{def_omega}), we start off by proving that $\psi$ is a solution of the Dirac equation. Let us assume that $\psi_1$ be a solution of the massless Dirac equation, then there exists a conserved current quantity j 1-form such that \cite{pollock}:

\begin{equation}
j_i = \bar{\psi_1} \gamma_i \psi_1.
\end{equation}

\noindent
If we prove that $S \psi_1$ is a solution of the massless Dirac equation (here S is a spinor transformation matrix such that $S \psi_1 = \psi$ ), then we have proven that eq. ($\ref{def_omega}$) holds with $\psi$ a solution to the massless Dirac equation. And here comes the proof of this. Let:

\begin{equation}
\slashed{D} S \psi_1 = \mathcal{O}.
\end{equation}

\noindent
We will seek a decomposition of the S matrix in the basis of the gamma matrices ($\gamma_{\mu_1} \cdots \gamma_{\mu_n}$). We find that the most difficult case is the case for a 4-dimensional manifold, and we write the decomposition in that case, leaving the case for higher dimensional manifolds and implicitly gamma matrices for the reader. So in 4 dimensions we choose as a basis {$I, \gamma_\mu, \sigma_{\mu \nu}, \gamma_5$}, where

\begin{equation}
\sigma_{\mu\nu} =-\frac{i}{2}(\gamma_\mu \gamma_\nu - \gamma_\nu \gamma_\mu).
\end{equation} 

\noindent
Note also that gamma matrices in curved spacetime can be written in terms of constant gamma matrices with flat indices as \cite{freedman}:

\begin{equation}
\gamma_\mu(x) = {e^a_\mu (x) \gamma_a}
\end{equation}

\noindent
So we can write the decomposition of the S matrix as:

\begin{equation}
S = \alpha^\mu (x)\gamma_\mu + \alpha(x) I +\beta^{\mu\nu} (x) \sigma_{\mu\nu} + \beta^5 (x)\gamma_5
\end{equation}

\noindent
where $ \alpha^\mu (x) = \alpha^a e^\mu_a(x)$, $ \alpha(x) = \alpha^a_\mu e^\mu_a (x)$, $\beta^5 (x) = \beta^a_\mu e^\mu_a (x)$ and 
$\beta^{\mu\nu} (x) = \beta^{ab} e^\mu_a (x) e^\nu_b (x)$. Here $\alpha^a , \alpha^a_\mu, \beta^a_\mu, \beta^{ab}$ are constants.

\medskip
\noindent
According to the vielbein postulate it follows that:

\begin{equation}
\nabla_\nu (\gamma_{\mu_1} \cdots \gamma_{\mu_n}) = 0 
\end{equation}

\noindent
and so then:

\begin{equation}
\slashed{D} (\gamma_{\mu_1} \cdots \gamma_{\mu_n}) = 0 .
\end{equation}

\noindent
It then follows that:

\begin{equation}
\mathcal{O}= \slashed{D} S \psi_1 = S \slashed{D} \psi_1 = 0
\end{equation}

\noindent
So $S\psi_1$ is a solution to the massless Dirac equation and then $\psi$ in equation (\ref{def_omega}) is a solution of the massless Dirac equation, too.

\section{Solution for the twistor equation}
\noindent
In this section we prove that $\psi$ is a solution to the twistor equation also, and that makes $\psi$ a Killing spinor.  The twistor spinor equation is \cite{C47}:

\begin{equation}\label{twistor_eq}
\nabla_X \psi = - \frac{1}{n} X \vee \slashed{D} \psi.
\end{equation}

\noindent
Please note that the Clifford product is defined as:

\begin{equation}
e_\mu  \vee = \gamma_\mu
\end{equation}

\noindent
and has the following property-

\begin{equation}\label{PropertyClifford}
\nabla_Y (X \vee \psi) = (\nabla_Y X) \vee \psi + X \vee (\nabla_Y \psi).
\end{equation}

\noindent
We start from the definition of a rank p conformal Killing-Yano tensor \cite{kubi}:

\begin{equation}
\nabla_X k= \frac{1}{p+1}X \biggl\lrcorner dk -\frac{1}{D-p+1}X^\flat \vee \delta k
\end{equation}

\noindent
In our case the CKY $\omega$ is closed and harmonic, so:
 
\begin{equation}
0=\nabla _X \omega_\mu = \nabla_X(\bar{\psi} \gamma_\mu \psi) = \nabla_X (\bar{\psi} e_\mu \vee \psi).
\end{equation}

\noindent
Expanding the expression and using the properties of the Clifford product in rel. ($\ref{PropertyClifford}$) and also the vielbein postulate, we get:

\begin{equation}
(\nabla_X \bar{\psi}) (e_\mu \vee \psi) + \bar{\psi} e_\mu \vee \nabla_X \psi = 0.
\end{equation}

\noindent
And so writing this equation as a scalar product \cite{semmelmann, cariglia}:

\begin{equation}\label{PropertyScalar}
<(\nabla_X \psi), (e_\mu \vee \psi)> = - < \psi, e_\mu \vee \nabla_X \psi>.
\end{equation}

\noindent
This is equivalent to:

\begin{equation}
<\psi, \nabla_X(e_\mu \vee \psi)> = -  < \psi, e_\mu \vee \nabla_X \psi>
\end{equation}

\noindent
and so again with the vielbein postulate:

\begin{equation}
<\psi, (e_\mu \vee \nabla_X \psi)> = -  < \psi, e_\mu \vee \nabla_X \psi>
\end{equation}

\noindent
and it follows that-

\begin{equation}
e_\mu \vee \nabla _X \psi = 0.
\end{equation}

\noindent
And taking the Clifford product with $e^\mu$:
\begin{equation}
e^\mu \vee e_\mu \vee \nabla _X \psi = 0.
\end{equation}

\noindent
And so, using the Dirac equation $(\slashed{D}-\mu) \psi=0$:

\begin{equation}
e^\mu \vee e_\mu  \vee \nabla_X \psi = - X \vee (\slashed{D}-\mu) \psi.
\end{equation}

\noindent
Hence with $\mu=0$, so a massless spin, and taking into account that $e^\mu \vee e_\mu \vee= n I_n$:

\begin{equation}
\nabla_X \psi  = -\frac{1}{n} X \vee \slashed{D} \psi.
\end{equation}

\noindent
And this is indeed the twistor equation. Naturally, other Killing spinors may exist, but the existence of the closed, harmonic CKY tensor guarantees that at least this Killing spinor corresponding to $\mu=0$ exists.

\section{Examples of non-compact manifolds with Killing spinors}

\medskip
\noindent
We present here a couple of examples of non-compact manifolds endowed with a harmonic, closed conformal Killing-Yano tensor and we prove that there exists a Killing spinor on it.

\medskip
\noindent
Our examples have the form of 8-dimensional manifolds with Spin(7) holonomy metrics \cite{Gary}:

\begin{multline}
ds^2_8 = \frac{(r+l)^2 dr^2}{(r+3l)(r-l)} + \frac{l^2 (r+3l)(r-l)}{(r+l)^2} \sigma^2 +
\frac{1}{4} (r+3l)(r-l) (D \mu^i)^2 + \\ \\
+\frac{1}{2} (r^2- l^2) d\Omega^2_4
\end{multline}

\noindent
which for r positive is topologically $\mathbb{R}^8$ and is called $\mathbb{A}_8$. Then for r negative, we get the 8-dimensional $\mathbb{B}_8$ manifold with topology $\mathbb{R}^4 X S^4$:

\begin{multline}
ds^2_8 = \frac{(r-\tilde{l})^2dr^2} {(r-3\tilde{l})(r+\tilde{l})} +\frac{\tilde{l}^2 (r-3 \tilde{l})(r+
	\tilde{l})}{(r-\tilde{l})^2} \sigma^2 +\frac{1}{4}(r-3\tilde{l})(r+\tilde{l})(D\mu^i)^2 +\\ \\
+\frac{1}{2}(r^2-\tilde{l}^2) d \Omega^2_4 
\end{multline} 

\noindent
with

\begin{equation}
\mu_1 = sin \theta sin \psi, \hspace{0.5cm} \mu_2= sin \theta  cos \psi, \hspace{0.5cm} \mu_3= cos \theta,     
\end{equation}

\noindent
and

\begin{equation}
D \mu^i = d \mu^i + \epsilon_{ijk} A^j_{(1)} \mu^k, \hspace{0.25cm} \sigma = d \phi + \mathcal{A}_{(1)}, \hspace{0.25cm}
\mathcal{A}_{(1)}= cos \theta d \psi - \mu^i A^i_{(1)},
\end{equation}

\noindent
and

\begin{multline}
\sum_i (D\mu^i)^2 = (d \theta - A^1_{(1)} cos \psi + A^2_{(1)} sin \psi)^2 + \\ \\
sin^2 \theta (d \psi+ A^1_{(1)} cot \theta sin \psi +A^2_{(1)} cot \theta cos\psi- A^3_{(1)} )^2.
\end{multline}

\noindent
and $A^i_{(1)}$ is the SU(2) Yang-Mills instanton on $S^4$. Since the manifolds have Spin(7) holonomy then it follows that there exists a covariant constant spinor \cite{Gary}:

\begin{equation}
\eta = e^{\frac{1}{2} \theta \Gamma_{71}} e^{\frac{1}{2} \psi \Gamma_{12}} \eta_0
\end{equation}

 and a closed, harmonic conformal Killing-Yano form can be constructed, such that
 
 \begin{equation}
 k = \bar{\eta} \Gamma_{AB} \eta, \hspace{1.0cm} \nabla_X k=0.
 \end{equation}

\medskip
\noindent
The 2-form has then the expression:

\begin{equation}
k = -\hat{e}^2 \wedge \hat{e}^7 - \hat{e}^5 \wedge \hat{e}^6 + \hat{e}^7 \wedge \hat{e}^8 +\frac{\partial \hat{Y}_{(2)}}{\partial \theta} +\frac{1}{sin \theta} \frac{\partial \hat{Y}_{(2)}}{\partial \psi} 
\end{equation}

\medskip
\noindent
where  $\hat{e}^\alpha = c e^\alpha$ and c is the solution of the following system of equations \cite{Gary}:

\begin{equation}
\dot{a} = 1 - \frac{b}{2a} -\frac{a^2}{c^2}, \hspace{0.25cm} \dot{b} = \frac{b^2}{2 a^2}- \frac{b^2}{c^2}, \hspace{0.25cm} \dot{c} =\frac{a}{c} +\frac{b}{2c}.
\end{equation}
where $(e^3,e^4, e^5, e^6)$ are the basis of tangent space 1-forms on the unit $S^4$ and the tangent space index 7 is in the radial direction and 8 in the U(1) fiber direction. Then 

\begin{equation}
\hat{Y}_{(2)} = \frac{1}{2}[sin \theta (cos \psi F^1_{\alpha \beta} + sin \psi F^2_{\alpha \beta}) + cos \theta F^3_{\alpha\beta} ] \hat{e}^\alpha \wedge \hat{e}^\beta,
\end{equation}

\noindent
and-

\begin{equation}
F^1_{(2)}= -(e^4 \wedge e^5 + e^3 \wedge e^6)
\end{equation}

\begin{equation}
F^2_{(2)} = -(e^5 \wedge e^3 + e^4 \wedge e^6)
\end{equation}

\begin{equation}
F^3_{(2)}= -(e^3 \wedge e^4 + e^5 \wedge e^6).
\end{equation}

\medskip
\noindent
The only thing left to prove is that $\eta$ is Killing spinor and so we start from:

\begin{equation}
\nabla_X k_{\mu \nu} = \nabla_X(\bar{\eta} \gamma_\mu \gamma_\nu \eta)=0
\end{equation}

\noindent
So

\begin{equation}
(\nabla_X \bar{\eta}) (e_\mu \vee e_\nu \vee \eta) + \bar{\eta} e_\mu \vee e_\nu \vee \nabla_X \eta = 0.
\end{equation}

\noindent
In scalar product:

\begin{equation}
<(\nabla_X \eta), (e_\mu \vee e_\nu \vee \eta)> = - < \eta, e_\mu \vee e_\nu \vee \nabla_X \eta>.
\end{equation}

\noindent
This is equivalent to:

\begin{equation}
<\eta, \nabla_X(e_\mu \vee e_\nu \vee \eta)> = -  < \eta, e_\mu \vee e_\nu \vee \nabla_X \eta>
\end{equation}

\noindent
and so again with the vielbein postulate it follows that :

\begin{equation}
e_\mu \vee e_\nu \vee \nabla _X \eta = 0.
\end{equation}

\noindent
Taking into account the Dirac equation, one gets the twistor equation:

\begin{equation}
\nabla_X \eta  = -\frac{1}{n} X \vee \slashed{D} \eta.
\end{equation}

\medskip
\noindent
We have then proven for manifolds $\mathbb{A}_8$ and $\mathbb{B}_8$ that a harmonic closed conformal Killing-Yano form exists and that it implies the existence of a covariant constant Killing spinor.

\section{Conclusions}

\medskip
\noindent
In this paper we generalize to non-compact manifolds a result known to be true only in the compact. We also provide examples of how our result works. It was proven in literature that for Sasaki-Einstein manifolds (which are compact spaces) Killing spinors exist, \cite{Kath}, and the Hijazi inequality result, \cite{lichnerowicz}, ensures that Killing spinors exist for compact spaces in general. By using the methods of differential geometry and relying on previous work done by Lichnerowicz and others, we prove and generalize this result to non-compact manifolds with invertible Dirac operator. We demonstrate that if a harmonic, closed conformal Killing-Yano tensor exists in this case, then a Killing spinor corresponding to a massless Dirac spinor exists as well. In that, we proved (sections 3 and 4) that a solution $\psi$ exists, and it is a Killing spinor, since it is both the solution of the Dirac equation and of the twistor spinor Eq. (\ref{twistor_eq}). By mathematical induction the same can be proved for an arbitrary p-form $\omega$ that is a harmonic, closed conformal Killing-Yano tensor.   

\medskip
\noindent
As directions for future work, it would be interesting to see if it can be establish a Lichnerowicz-type result for certain complete or non-compact manifolds, and to find out what conditions should be met such that the Dirac operator is invertible in these cases. Even though a universal Hijazi inequality is difficult the be found for non-compact spaces, it would be interesting to find the conditions under which such a bound generating Killing spinors exists for non-compact spaces. Since constructing symmetry operators for the Dirac equations on various types of manifolds was proved to be useful to further classify manifolds, and to study their hidden symmetries, in our work we aim to study the symmetry operators associated with Killing spinors for the same manifold and thus to discover new structures.

\section*{Acknowledgements}
	The authors thank Virgil Baran, Gary Gibbons, George Papadopoulos and Claude Warnick for reading the manuscript and making useful comments and CR acknowledges the generous support of a POS-DRU European fellowship.

\end{document}